\documentclass[a4paper,10pt]{article}

\title{  Incorporation of Generalized Uncertainty Principle 
into Lifshitz   Field Theories}
\author{Mir Faizal$^1$ and Barun Majumder$^2$\\
$^1$Department of Physics and Astronomy, \\  University of Waterloo,   Waterloo,\\
Ontario N2L 3G1, Canada \\
f2mir@uwaterloo.ca \\
$^2$ Indian Institute of Technology Gandhinagar,\\
Ahmedabad,  382424, India\\
barunbasanta@iitgn.ac.in }
\date{}
\begin{document}

\maketitle

\begin{abstract}
In this paper, we will   incorporation the generalized uncertainty principle
into field theories with Lifshitz scaling. 
We will first construct  both bososnic and fermionic  theories with Lifshitz scaling based on  
generalized uncertainty principle. 
After that we will incorporate the generalized uncertainty principle into an 
 non-abelian gauge theory with Lifshitz scaling.  We will observe that even though the action for this theory 
is non-local, it is invariant under local gauge transformations. 
We will also perform the  stochastic quantization of this Lifshitz  fermionic  theory based  
generalized uncertainty principle. 
\end{abstract}
\section{Introduction}
 The classical picture of spacetime   breaks down in most approaches to quantum gravity.
 This is due to the fluctuations in the geometry being of order one at Planck scale. Thus, the picture of
 spacetime as a continuous differential manifold cannot be valid below Planck length. Furthermore, the existence of a minimum length 
 scale is also a feature of string 
 theory \cite{unz2}-\cite{un2z}. 
In fact,  in loop quantum gravity  
the existence of minimum length 
 turns big bang into a big bounce \cite{unz1}. However, the existence of minimum length is not consistent with conventional 
 uncertainty principle, which states that one can measure length with arbitrary accuracy, if one takes no measurement of momentum 
  \cite{un1}-\cite{un54}.  
 Thus, the uncertainty principle has to be modified if one wants to incorporate the existence of minimum length scale. 
 These considerations have led to  a  modification of the Heisenberg uncertainty principle, which in turn has led to a 
 modification of the Heisenberg algebra. 
It may be noted that the implications of this modified 
uncertainty principle for quantum field theory have also been studied \cite{unn9}-\cite{unn}. 
In this paper, we analyse a quantum field theory based on generalized uncertainty with Lifshitz scaling. 
Lifshitz field theories are quantum field theories based on an anisotropic scaling between space and time. 

 Lifshitz
theories were first introduced in
condensed matter physics to model quantum criticality \cite{1}-\cite{4}. In fact, 
a Fermi-surface-changing Lifshitz transition occurs for some heavy fermion compounds \cite{15}.
The location
of this Fermi-surface-changing Lifshitz transition  is influenced by 
carrier doping. Due to strong correlations, a heavy band does not shift rigidly with 
the chemical potential and
the actual shift is determined by the interplay of heavy and additional light bands 
crossing the Fermi level. 
Furthermore, 
meta-magnetic transitions in models for heavy fermions has also been analysed 
using doped Kondo lattice model in two dimensions \cite{16}.
Some  heavy fermion metals displays a field-driven quantum phase transition due to a  
breakdown of the
Kondo effect \cite{a1}-\cite{a12}. Many  of the properties  have been  described
by a Zeeman-driven Lifshitz transition of narrow heavy fermion bands \cite{a2}. 
 Materials that  cannot
be described with the local dielectric response have been described by a generalization of the usual Lifshitz theory
\cite{a3}. In fact,  the temperature correction to the Casimir-Lifshitz 
free energy between two parallel plates made of dielectric 
material, possessing a constant conductivity at low temperatures, has  been calculated \cite{a4}. 
Lifshitz theory have also been  used for calculating 
the van der Waals and Casimir interaction between graphene 
and a material plate, graphene and an atom or a molecule, and between a single-wall carbon nanotube and a 
plate \cite{a5}. 
In this model the reflection properties of electromagnetic oscillations on graphene are governed by the specific 
boundary conditions imposed 
on the infinitely thin positively charged plasma sheet, carrying a continuous fluid with some mass and charge density. 

Fermionic retarded Green's function  with  $z = 2$ has been studied 
at finite temperature and finite chemical potential \cite{17}. 
Here the usual Lifshitz geometry was replaced by a Lifshitz black hole. 
 Hawking radiation for Lifshitz fermions has also been studied \cite{18}. 
 Fermionic theories  with $z=2$ Lifshitz scaling have also been constructed using  a  
non-local differential operator \cite{6a}. 
This  non-local differential operator is defined 
 using   harmonic extension of a compactly 
 supported function  \cite{7a}-\cite{12a}. It appears as a map from the  Dirichlet-type problem to the 
 Neumann type problem. 
It may be noted that 
fermionic theories with $ z = 3$ have also been studied  \cite{2a}-\cite{3a}. 
 It has been demonstrated that 
 Nambu-Jona-Lasinio type four-fermion coupling at the $z=3$ Lifshitz  fixed point in four
 dimensions is asymptotically free and generates a mass scale  \cite{5a}. 
In this paper, we will study both bososnic and fermionic Lifshitz field theory, 
consistent with generalized uncertainty principle. We will also study the gauge symmetry for these theories. 
It may also be noted that another interesting deformation of quantum mechanics comes from 
 stochastic quantization  \cite{stochastic}-\cite{stochastic1}. 
 Stochastic quantization has provided a powerful framework for analysing bosonic theories with 
 Lifshitz scaling \cite{bh}-\cite{b11h}. 
 In fact, effect of ohmic noise on the non-Markovian spin dynamics resulting in Kondo-type correlations
 have been studied using  stochastic quantization  \cite{ko}
 In this paper, we will analyse the    stochastic quantization
   of Lifshitz Dirac equation with minimum length.

\section{Generalized Uncertainty Principle}

In the Lifshitz field theories the
  scaling is usually taken as $x \to bx$ and $t \to b^z t$, where 
$b$ is called the scaling factor and 
$z$ is called the degree of anisotropy. For $z =1$, this reduces to the usual 
conformal transformation. In this paper, we will analyze the Lifshitz theories with 
$z =2$. 
The Lifshitz action for a  bosonic field  with $z=2$,  can be written as \cite{6a}
\begin{eqnarray}
S_{b}&=& \frac{1}{2}\int d^{d+1}x~(  \phi \partial^0 \partial _{0}\phi
- \kappa ^{2}\phi  (\partial^i \partial_i )^2 \phi).
\end{eqnarray} 
The Lifshitz theories are unitarity becuase they contain no   higher order   temporal derivatives. 
So, we will leave the temporal part of the Lifshitz action for a  bosonic field undeformed. However, we will
deform its spatial part, to make it consistent with the 
existence of a minimum measurable length \cite{un54}-\cite{unn9}. 
The  
Heisenberg uncertainty principle is not consistent with the existence of a minimum measurable length, as 
according to it, we can measure length up to arbitrary accuracy, if we do not measure the momentum. So, to 
accommodate the existence of a minimum measurable length scale, the Heisenberg uncertainty principle has to be modified to 
the generalized uncertainty principle. The generalized uncertainty principle can 
derived from a deformed  Heisenberg algebra.
The deformation  of the Heisenberg algebra in turn deforms the coordinate representation of the momentum 
operator, and this deforms  the 
Laplacian to $ \partial^i\partial_i \to \partial^i\partial_i (1- \beta \partial^j \partial_j )$ \cite{unn9}. 
Now  using this definition of the deforms  the 
Laplacian, the deformed Lifshitz action  can be written as  
\begin{eqnarray}
S_{b}&=& \frac{1}{2}\int d^{d+1}x ~\left(  \phi \partial^0 \partial _{0}\phi
- \kappa ^{2}\phi  [ \partial^i \partial_i  (1 - \beta \partial^j \partial_j) ]^2 \phi \right).
\end{eqnarray}
Here we have to promote  that parameter $\beta$ to a background field, such that it scales as $\beta \to b^2 \beta $.
This ensures that theory still has Lifshitz scaling
 after it has been deformed by the generalized uncertainty principle.
It may be noted that  
it is common to promote parameters in conformal field theories to background field in this way \cite{aqwe}-\cite{aqwer}. 
These background fields have scaling properties that ensures the conformal invariance of the deformed theory. 
  Now we can write this action as
\begin{eqnarray}
S_{b}&=&  \frac{1}{2}\int d^{d+1}x~\left(
\phi \partial^0 \partial _{0}\phi- \kappa ^{2} \partial ^{i}\phi \mathcal{T}^2_{\partial}  (1 - \beta \partial^j \partial_j)^2 
 \partial _{i}\phi \right), 
\end{eqnarray} 
where $\mathcal{T}_{\partial}= \sqrt{- \partial^i \partial_i} $. 
It may be noted that the   non-local differential operator
$\mathcal{T}_{\partial}  $  is crucial in constructing the 
 fermionic action with  Lifshitz scaling.

Even though this operator is non-local it  can be  effectively viewed as a 
local operator, by using the theory of  harmonic extension of  functions from 
$R^d$ to $R^d \times (0, \infty)$ \cite{6a}-\cite{12a}. 
Thus, we can define $\mathcal{T}_{\partial}$ by its action on 
 functions 
 $f: R^d \to R $, such that its  harmonic extension $u: R^d \times (0, \infty) \to R$ satisfies, 
$
\mathcal{T}_{\partial} f(x) = -\partial_y u (x, y)| _{y =0}
$. 
This is because if we start with a function $f: R^d \to R$, and  
find a harmonic function $u: R^d \times (0, \infty) \to  R$, such that
 its restriction 
to $R^d$ coincides with the original function $f: R^d \to R$, then  
 it is possible to 
find $u$ by solving 
a Dirichlet problem. This  Dirichlet problem can be expressed in terms of 
 the Laplacian in $R^{d+1}$, which is denoted by $\partial^2_{d+1}$. So, for $x \in R^d$ and $ y \in R$, 
we have,  
$u(x, 0) = \phi (x)$ and $ \partial^2_{d+1} u (x, y) =0  $.  
In fact,  for a smooth function $C^\infty_0 (R^d) $, there is a unique  harmonic extension 
 $ u \in C^\infty (R^d \times (0, \infty))$. 
Now as  $\mathcal{T}_{\partial} \phi (x)$ also has a harmonic
extension to $R^d \times (0, \infty)$, we can obtain the following result, $
\mathcal{T}_{\partial}^2 \phi (x) = \partial^2_y u(x, y)|_{y =0}
= - \partial^i \partial_i u(x, y)|_{y =0}$. 
Thus, it is possible to define  $\mathcal{T}_{\partial} = \sqrt{- \partial^i \partial_i}$,   
because $\mathcal{T}_{\partial}^2 \phi(x) = - \partial^i \partial_i  \phi (x)$. 
So, we can write $\mathcal{T}_{\partial} \exp ikx  = |k| \exp ikx$, because 
$\mathcal{T}_{\partial}^2 \exp ikx  = |k|^2  \exp ikx$. 
Furthermore, if  we start with two fields $\phi_1 (x)$ and $\phi_2 (x)$,  such that 
$u_1(x, y)$ and $u_2 (x, y)$ are their 
 harmonic extensions to $ C = R^d \times (0, \infty)$, and 
 both   these harmonic extensions vanish 
for $|x| \to \infty $ and $|y| \to \infty $, then we can write  \cite{5a01}
\begin{equation}
\int_C d^d x dy~ u_1(x, y) \partial_{n+1}^2 u_2 (x, y)  - 
\int_C d^d x dy~ u_2(x, y) \partial_{n+1}^2 u_1 (x, y) 
= 0. 
\end{equation}
Thus, we get the following expression 
 \begin{equation}
 \int_{R^d} d^d x~ \left(u_1(x, y) \partial_y u_2 (x, y)  -   u_2(x, y) 
\partial_x u_1 (x, y) \right)\left. \right|_{y =0}  
= 0. 
 \end{equation}
This  can now be written in terms of $\phi_1 (x) $ and $\phi_2 (x)$ as 
\begin{equation}
 \int_{R^d}  d^dx~ \left(\phi_1(x) \partial_y
 \phi_2 (x) -  \phi_2(x)\partial_x \phi_1 (x)\right) 
= 0. 
 \end{equation}
So, the operator $\mathcal{T}_{\partial}$ can be moved from $\phi_2 (x)$ to $\phi_1 (x)$,  
 \begin{equation}
 \int_{R^d}d^d x~ \phi_1 (x) \mathcal{T}_{\partial} f\phi_2 (x)   =  
 \int_{R^d}  d^d x ~\phi_2 (x) \mathcal{T}_{\partial} \phi_1 (x).
 \end{equation}

Now the Lifshitz bososnic  action, consistent with generalized uncertainty principle,  can also be written as 
\begin{equation}
S_b =\frac{1}{2}\int d^{d+1}x~\partial ^{\mu }\phi ~G_{\mu \nu }
\partial ^{\nu }\phi,
\end{equation}
where $G_{\mu\nu}$ can be written as 
\[  \begin{array}{ccc}
G_{\mu\nu}
 \end{array}  =  \left( \begin{array}{ccc}
I_{1\times1} & 0_{1\times d }  \\
0_{d \times 1 } & - \kappa^2 \mathcal{T}^2_{\partial} (1 - \beta \partial^j \partial_j)^2 I_{d \times d} \\
 \end{array} \right). \]
This equation can now be regarded as defining a scalar product 
for vector fields, such that for any two vectors $V$ and $W$, we have 
\begin{eqnarray}
 (V(x), W(x)) = \int d^{d+1}x~ \left(V_0 W_0 - \kappa^2 V_i\mathcal{T}^2_{\partial} (1 - \beta \partial^j \partial_j) W_i \right). 
\end{eqnarray}
The under group of isometries this inner product remains invariant. So, we can write,  $(\Lambda (V), \Lambda (W)) = (V, W)$. Thus, 
we can write $\Lambda_{0\mu} \Lambda_{0\nu} - \kappa^2 \Lambda_{i\mu} \Lambda_{i\nu} \mathcal{T}_\partial (1 - \beta \partial^j \partial_j) =
G_{\mu\nu}$. From this we can infer that $\Lambda_{00} = \Lambda_{i0} = \Lambda_{0i} =0$ and 
$\Lambda^k_{i}\Lambda_{kj} = \delta_{ij}$.
Now a set of local gamma matrices can be defined, such that 
$
  \{ \Gamma_\mu, \Gamma_\nu\} = 2 G_{\mu\nu}
$. Furthermore, an appropriate choice for these 
local gamma matrices is $\Gamma_0 = \gamma_0$ and $\Gamma_i = \kappa \mathcal{T}_{\partial} 
(1 - \beta \partial^j \partial_j) \gamma_i$, where 
$
  \{ \gamma_a, \gamma_b\} = 2 \eta_{ab}
$. 
We can thus define a fermionic Lifshitz  operator as 
$\Gamma^\mu \partial_\mu = 
 \gamma^0 \partial _{0 } +  \kappa\gamma^i  \mathcal{T}_{\partial} (1 - \beta \partial^j \partial_j)  \partial _{i } $. 
We observe that $ \Gamma^\mu \partial _{\mu }
 \Gamma^\nu \partial _{\nu } = \partial^0\partial_0 - \kappa^2 [\partial^i\partial_i (1 - 2 \beta \partial^j p\partial_j)]^2$.  
The Lifshitz action for a massless fermionic field can be written as  
\begin{eqnarray}
S_f&=& \frac{1}{2}
\int d^{d+1}x~\bar \psi\left( \Gamma^\mu \partial _{\mu }\right) \psi \nonumber \\ 
&=&\frac{1}{2}
\int d^{d+1}x~ \bar \psi \left( \gamma^0 \partial _{0 } + \gamma^i \kappa \mathcal{T}_{\partial}(1 - \beta \partial^j \partial_j)  \partial _{i }\right) \psi. 
\end{eqnarray}

\section{ Gauge Symmetry}
In this section, we will analyse gauge theories with Lifshitz corresponding to generalized uncertainty principle. 
We note that if the   covariant derivative is gauge covariant, then so,  is  any function of the covariant derivative.
We will construct a covariant derivative from using a non-abelian gauge field $A_\mu = A_\mu^A T_A $, where 
$  [T_A, T_B] = i f_{AB}^C T_C $. Now if $\psi \to U  \psi$, then we should have $ D_\mu \psi \to U D_\mu \psi$.  
We can construct a covariant derivative with this transformation property if, 
we assume that the gauge field transforms as $A_\mu \to iU D_\mu U^{-1} $ and define the 
gauge covariant derivative as,  $D_\mu = \partial_\mu + i A_\mu $. This is because now the covariant derivative will transform as 
\begin{equation}
D_\mu \to U D_\mu U^{-1},
\end{equation}
 and so,  $ D_\mu \psi \to U D_\mu \psi$,  if $\psi \to U  \psi$. 
 Now any function of the covariant derivative is also gauge covariant. So, if we take a general function of $D_\mu $, 
 $  f(D^\nu D_\nu)  D_\mu $, then it transforms as 
 \begin{equation}
  f(D^\nu D_\nu)  D_\mu \to U f(D^\nu D_\nu)  D_\mu  U^{-1}, 
 \end{equation}
such that, $f(D^\nu D_\nu)  
D_\mu \psi \to U f(D^\nu D_\nu)  D_\mu \psi $. 

We can now use different $f (D^\nu D_\nu)   $, for the spatial and temporal part of the covariant derivative. 
Now we define $f_1(D^\nu D_\nu)  $ to be the function for the temporal  part of the derivative and 
$f_2  (D^\nu D_\nu)   $ to be the function for the spatial part of the derivative. 
The theory has  Lifshitz scaling, if we choose 
\begin{eqnarray}
 f_1(D^\nu D_\nu)D_0 &=& D_0, \nonumber \\
   f_2(D^\nu D_\nu) D_i &=&\kappa \mathcal{T}_{D}D_i , 
\end{eqnarray}
where $\mathcal{T}_{D}= \sqrt{- D^i D_i}$. 
 The covariant derivative  will still transform as
\begin{eqnarray}
 D_0  &\to & U D_0 U^{-1}, \nonumber \\
 \kappa \mathcal{T}_{D} D_i &\to & U \kappa \mathcal{T}_{D}D_i U^{-1}. 
\end{eqnarray}
Here 
we have to again   assumed that $\beta$ is  background field which scales like $\beta \to b^2 \beta $.
\cite{aqwe}-\cite{aqwer}. This ensures that the   theory constructed also has Lifshitz scaling. 
However, we also want a theory that will correspond to generalized uncertainty principle. In particular the matter part of the Lagrangian 
should reduce to the Lagrangian derived in the previous section, if we set all the gauge field to zero. 
Thus, we re-define $f_1(D^\nu D_\nu)$ and $f_2(D^\nu D_\nu)$ as 
\begin{eqnarray}
 f_1(D^\nu D_\nu)D_0 &=& D_0, \nonumber \\
   f_2(D^\nu D_\nu) D_i &=& \kappa \mathcal{T}_{D}(1 - \beta D^j D_j)  D _{i }. 
\end{eqnarray}
It may be noted that the  covariant derivative will still transforms as 
\begin{eqnarray}
 D_0  &\to & U D_0 U^{-1}, \nonumber \\
  \kappa \mathcal{T}_{D}(1 - \beta D^j D_j)  D _{i } &\to & U \kappa \mathcal{T}_{D}(1 - \beta D^j D_j)  D _{i } U^{-1}. 
\end{eqnarray}
So, we can now write the final action as 
\begin{equation}
 S 
=\frac{1}{2}
\int d^{d+1}x~ Tr [ \bar \psi ( \gamma^0 D_0
+ \gamma^i \kappa \mathcal{T}_{D}(1 - \beta D^j D_j)  D _{i }) \psi ].
\end{equation}
 Now the  temporal part of this action is invariant under local gauge transforms  because,  
$Tr [ \bar \psi ( \gamma^0 D_0
 ) \psi ] \to Tr [ \bar \psi U^{-1} U( \gamma^0   D_0  
 )   U^{-1} U \psi ] = Tr [ \bar \psi ( \gamma^0 D_0
 ) \psi ]$,  and the spatial part of this action is also  invariant under local gauge transforms because,   $Tr [ \bar \psi (  
 \gamma^i \kappa \mathcal{T}_{D}(1 - \beta D^j D_j)  D _{i }) \psi ] \to Tr [ \bar \psi U^{-1} U(   
  \gamma^i \kappa \mathcal{T}_{D}(1 - \beta D^j D_j)  D _{i })U^{-1} U \psi ] = Tr [ \bar \psi (  
  \gamma^i \kappa \mathcal{T}_{D}(1 - \beta D^j D_j)  D _{i }) \psi ]$. So, even though this action is non-local, it 
  is invariant under local gauge transformations, $A_\mu \to i U  D_\mu U^{-1}$.

It may be noted that we can now define a gauge field tensor for this theory as 
\begin{eqnarray}
 F_{i0} &=& -i [D_0,  \kappa \mathcal{T}_{D}(1 - \beta D^j D_j)  D _{i }], \nonumber \\
 F_{ij }&=& -i [ \kappa \mathcal{T}_{D}(1 - \beta D^k D_k)  D _{i }, \kappa \mathcal{T}_{D}(1 - \beta D^l D_l)  D _{j} ].  
\end{eqnarray}
It transforms as 
\begin{eqnarray}
F_{i0} &\to&  -i  [ UD_0 U^{-1},  U \kappa \mathcal{T}_{D}(1 - \beta D^j D_j)  D _{i }U^{-1}] , \nonumber \\
 &&= U F_{i0} U^{-1}, \nonumber \\
 F_{ij } &\to& -i [  U\kappa \mathcal{T}_{D} U^{-1}(1 - \beta  UD^k U^{-1} U  D_k U{-1})  U D _{i } U^{-1}, \nonumber \\ && 
 U \kappa  \mathcal{T}_{D} U^{-1}(1 - \beta UD^l U^{-1} U D_l) U^{-1} U D _{j} U^{-1}  ]\nonumber \\
 && = U F_{ij } U^{-1}.  
\end{eqnarray}
Now we can write the action for the gauge part of the action  as follows, 
\begin{equation}
S_g =- \frac{1}{4} \int d^{d+1} x ~Tr [F^{\mu\nu} F_{\mu\nu}]. 
\end{equation}
It may be noted that even thought this action is non-local, it is invariant under local gauge transformations, 
$A_\mu \to i U  D_\mu U^{-1}$,  because, $Tr [F^{\mu\nu} F_{\mu\nu}] \to Tr [ UF^{\mu\nu}U^{-1}U F_{\mu\nu} U^{-1}]
= Tr [F^{\mu\nu} F_{\mu\nu}]$. 
Now we can  write the gauge fixing term for this theory,  
\begin{equation}
S_{gh} = \int d^{d+1} x ~ Tr[b \partial^0 A_0 - b \kappa  \partial^i \mathcal{T}_\partial (1 - \beta \partial^j \partial_j ) A_i].   
\end{equation}
The ghost term for corresponding to this gauge fixing term, can be written as 
\begin{equation}
 S_{gf} = \int d^{d+1} x ~Tr [\bar c\partial^0 D_0 c -  \kappa^2 \bar c \partial^i \mathcal{T}_\partial (1 - \beta \partial^j \partial_j )
  D_i (1 - \beta D^k D_k)\mathcal{T}_{D} c].  
\end{equation}

  \section{Stochastic Quantization}
In this section,   the stochastic quantization of the Lifshitz  fermionic  theory based  
generalized uncertainty principle will be analysed. To perform this analysis an 
 an extra fictitious time variable $\tau$ will be  introduced, such that 
 $\psi (x) \to \psi (x, t), $ and $ \bar\psi (x) \to \bar\psi (x, t) $ 
 \cite{stochastic}-\cite{stochastic1}. 
 We will also 
use an appropriate Kernel $K (x, y)$ to ensure the relaxation 
 process is such that the systems will approach equilibrium as $\tau \to 0$. 
The   anticommuting fermionic Gaussian noise,  $\eta (x, \tau)$ and $\eta  (x, \tau)$, will satisfy 
 \begin{eqnarray}
   \langle \eta(x, \tau) \rangle &=& 0 
 \nonumber \\ \langle \bar \eta (x, \tau) \rangle &=& 0, \\ \nonumber
  \langle \eta  (x', \tau')  \bar \eta (x, \tau) \rangle &=& 2 \delta(\tau - \tau' ) K(x, x') \delta^4 (x - x'). 
 \end{eqnarray}
The Langevin equations for the  Lifshitz  fermionic  theory  based on the generalized uncertainty principle can be written as 
 \begin{eqnarray}
  \frac{\partial \psi (x, \tau) }{\partial \tau } &=& - \int d^4 y K(x, y)  \frac{\delta S  [\psi, \bar \psi ]}{ \delta \bar \psi } + \eta (x, \tau),
  \\ \nonumber
    \frac{\partial \bar \psi (x, \tau) }{\partial \tau } &=&  \int d^4 y \frac{\delta S  [\psi, \bar \psi ]}{ \delta   \psi }K(x, y)
    + \bar \eta (x, \tau). 
 \end{eqnarray}
Here the action for the  Lifshitz  fermionic  theory based  
generalized uncertainty principle is given by 
\begin{equation}
 S 
=\frac{1}{2}
\int d^{d+1}x~ Tr [ \bar \psi ( \gamma^0 D_0
+ \gamma^i \kappa \mathcal{T}_{D}(1 - \beta D^j D_j)  D _{i }) \psi ].
\end{equation}
It may be noted that only $\psi$ and $\bar \psi $ are the  dynamical variables,  
as we are analysing the system on a fixed background. 
The  partition function for the Lifshitz  fermionic  theory based  
generalized uncertainty principle, can be expressed as 
 \begin{equation}
  Z = \int D \bar \eta D \eta  \exp \left( - \frac{1}{2} \int d^4 x d^4 y d\tau \bar \eta (x, \tau) K^{-1} (x, y ) \eta (y, \tau) \right).    
 \end{equation}
This partition function can now be expressed as 
\begin{eqnarray}
 Z &=& \int D \bar \psi D \psi  \det\left[\frac{\delta \bar \eta}{\delta \bar\psi}\right]^{-1}  
 \det\left[\frac{\delta  \eta}{\delta \psi}\right]^{-1} \nonumber \\ && \times 
  \exp \left( - \frac{1}{2} \int d^4 x d^4 y d\tau \bar \eta (x, \tau) K^{-1} (x, y ) \eta (y, \tau) \right). 
\end{eqnarray}
Finally,   using Langevin equations, we obtain, 
\begin{eqnarray}
 Z &=&  \int D \bar \psi D \psi  \det
 \left[K^{-1}\frac{\partial }{\partial \tau} - \frac{\delta^2 S[\bar\psi, \psi]}{ \delta \bar \psi \delta \psi } 
\right]^{-1}  
 \det \left[K^{-1}\frac{\partial }{\partial \tau} +  \frac{\delta^2 S[\bar\psi, \psi]}{ \delta  \psi \delta \bar \psi } 
\right]^{-1}\nonumber \\ && \times 
 \exp \left(- \frac{1}{2} \int d^4 x d^4 y d\tau \left[\frac{\partial \bar \psi}{\partial \tau} K^{-1} -
 \frac{\delta  S[\bar\psi, \psi]}{   \delta  \psi }  \right]\right. \nonumber \\ && \left. \times 
\left[\frac{\partial \psi }{\partial \tau}  +  K  \frac{\delta  S[\bar\psi, \psi]}{  \delta \bar \psi }  \right]
\right). 
\end{eqnarray}
Here  the determinant is defined as the regularized product of eigenvalues. 
This is done by using    ghost fields $ (c_1,\bar c_1 ,  c_2, \bar c_2 )$,  and writing  \cite{11st}, 
\begin{eqnarray}
\bar c_1   \det \left[ K^{-1} \frac{\partial}{\partial \tau} - \frac{\delta^2 S}{\delta \psi \delta \bar \psi } \right] 
c_1 = \lambda \bar c_1 c_1,  \nonumber \\ 
\bar c_2 \det \left[ K^{-1} \frac{\partial}{\partial \tau} - \frac{\delta^2 S}{\delta \psi \delta \bar \psi } \right]  
c_2 = \lambda \bar c_2 c_2. 
\end{eqnarray}
So, we get 
\begin{eqnarray}
 \mathcal{L}_{eff} &=& \frac{1}{2}   \frac{\partial \bar \psi}{\partial \tau} K^{-1}\frac{\partial  \psi}{\partial \tau} -
 \frac{1}{2}  \frac{\delta  S[\bar\psi, \psi]}{   \delta  \psi } K  \frac{\delta S[\bar\psi, \psi]}{   \delta  \bar \psi }
\nonumber \\  && +  \bar c_1 
 \left[K^{-1}\frac{\partial }{\partial \tau} - \frac{\delta^2 S[\bar\psi, \psi]}{ \delta \bar \psi \delta \psi } 
\right]c_1  \nonumber \\  &&   + 
 \bar c_2  \left[K^{-1}\frac{\partial }{\partial \tau} +  \frac{\delta^2 S[\bar\psi, \psi]}{ \delta  \psi \delta \bar \psi } 
\right] c_2. 
\end{eqnarray}
The auxiliary fields $\bar F$ and $F$ are introduced to  write the partition function as \cite{12st}-\cite{10st}
\begin{equation}
 Z = \int D\bar \psi D \psi Dc_1 D\bar c_1 Dc_2 D\bar c_2  D F D \bar F \exp\left( - \int d^4 x e  \mathcal{L}_{eff} \right), 
\end{equation}
where  
\begin{eqnarray}
 \mathcal{L}_{eff} &=&  2 \bar F K^{-1} F + i 
 \left[\frac{\partial \bar \psi }{\partial \tau} K^{-1} - \frac{\delta  S[\bar\psi, \psi]}{ \delta \psi } 
\right] F \nonumber \\ &&
+  i 
\bar F  \left[\frac{\partial   \psi }{\partial \tau}   + K \frac{\delta  S[\bar\psi, \psi]}{ \delta \bar \psi } 
\right] +  \bar c_1 
 \left[K^{-1}\frac{\partial }{\partial \tau} - \frac{\delta^2 S[\bar\psi, \psi]}{ \delta \bar \psi \delta \psi } 
\right]c_1  \nonumber \\  &&   + 
 \bar c_2  \left[K^{-1}\frac{\partial }{\partial \tau} +  \frac{\delta^2 S[\bar\psi, \psi]}{ \delta  \psi \delta \bar \psi } 
\right] c_2. 
\end{eqnarray}

This action can be written using the superfield formalism. Thus, complex superfields 
 superfields $\Omega (x, \tau, \theta, \bar \theta) $ and $\Omega (x, \tau, \theta, \bar \theta) $, are defined as 
\begin{eqnarray}
 \Omega(x, \tau, \theta, \bar \theta)  &=& \psi (x, \tau) + \bar \theta c_1(x, \tau)  + \bar c_2(x, \tau)  \theta + i \theta \bar \theta F(x, \tau) ,
 \nonumber \\
  \bar \Omega(x, \tau, \theta, \bar \theta)  &=& \bar \psi(x, \tau)  + \bar \theta c_2(x, \tau)  + \bar c_1(x, \tau)  \theta + i \bar \theta
  \theta \bar F(x, \tau) .
\end{eqnarray}
The following superderivatives and supercharges are also defined, 
\begin{eqnarray}
 D =  \frac{\partial }{\partial \bar \theta } - \theta \frac{\partial}{\partial \tau}, && Q =  \frac{\partial }{\partial \bar \theta }, 
 \nonumber \\
 \bar Q =  \frac{\partial}{\partial \theta} + \bar  \theta \frac{\partial}{\partial \tau}, &&\bar D = \frac{\partial}{\partial \theta}.  
\end{eqnarray}
The commutators   of these superderivatives and supercharges are given by 
\begin{eqnarray}
 \{ D, \bar D \} =  - \frac{\partial }{\partial  \tau}, &&
  \{ Q, \bar Q \} =  \frac{\partial }{\partial  \tau}. 
\end{eqnarray}
Now we can define an superspace action $ S = S_1 + S_2 + S_3$, where 
\begin{eqnarray}
 S_1 &=&  \int d^4 x e d\bar \theta d \theta \bar D \bar \Omega  D \Omega
 \\ \nonumber &=& \int d^4 xe  \left[ \bar c_1\frac{\partial c_1 }{\partial \tau}  + \bar F F + i \bar F \frac{\partial \psi}{\partial \tau}\right],
\\ \nonumber 
   S_2 &=&  \int d^4 x e d\bar \theta d \theta \bar D  \Omega  D \bar \Omega 
 \\ \nonumber &=& \int d^4 x e \left[ \bar c_2\frac{\partial c_2 }{\partial \tau}  +  F \bar F + i \bar F \frac{\partial \bar \psi}{\partial \tau}\right],
 \\ \nonumber 
 S_3 &=&  \int d^4 x e d\bar \theta d \theta S [ \bar \Omega, \Omega ]
 \\ \nonumber &=& \int d^4 x e \left[ \bar F \frac{\delta S_3}{\delta \bar \psi } +  \frac{\delta S_3}{\delta   \psi } F
 +   \bar c_1  \frac{\delta^2  S_3}{\delta \psi \delta \bar \psi } c_1  + \bar c_2 \frac{\delta^2  S_3}{\delta \psi  \delta \bar \psi  } c_2  \right]. 
\end{eqnarray}
It may now be noted that this  action $S$ coincides with the component action, if we make the following changes of variables 
$ \bar F  K^{-1}\to \bar F, 
 \bar  \psi K^{-1} \to \bar \psi, \bar c_1 K^{-1} \to \bar c_1, \bar c_2 K^{-1} \to \bar c_2 $,  and 
 $F \to F, \psi \to \psi, c_1 \to c_1, c_2 \to c_2$ \cite{ferm}. 

Thus,  we have performed stochastic quantization of the Lifshitz  fermionic  theory based  
generalized uncertainty principle using superspace formalism.  
It may be noted that stochastic quantization of this system induces a supersymmetry corresponding to the 
 extra fictitious time variable.

\section{Conclusion}
In this paper we deformed the Lifshitz field theories to make them consistent with the existence of a minimum measurable length. 
This was done by incorporating   generalized uncertainty principle in them. 
We had to promote a parameter used in the theory to a background field with interesting scaling properties, to preserve 
the Lifshitz scaling of the deformed theory. We also analysed a deformed Lifshitz theory  
gauge theory based on the  generalized uncertainty principle. We observed that even though this theory is non-local, it is invariant 
under local gauge transformations. We are expect to obtain similar results, if we generalize this work by  incorporating  
terms linear in the momentum,  in the deformed Heisenberg algebra \cite{line1}-\cite{l2}. It would also be interesting to 
analyse the BRST symmetry for this theory. We also performed the  stochastic quantization of the deformed Dirac equation 
semi-classically. It may be noted that the a deformed version of the general relativity has been obtained based on generalized 
uncertainty principle \cite{unn9}. It would be interesting to analyse  the effect a combination of Lipschitz scaling 
and generalized uncertainty principle can have on general relativity.  

The holographic dual to the Lifshitz field theory has also been studied \cite{13a}-\cite{14}. 
The   dual of the field theory vacuum has a bulk metric, 
\begin{equation}
 ds^2 = -r^{2z} dt^2 + r^2 dx^2 + L^2 r^{-2} dr^2, 
\end{equation}
where $L^2$  represents the overall curvature scale. It is obvious that for $z =1$, this metric reduces to the 
usual $AdS$ metric.  
In these Lifshitz theories, the 
 renormalization group flow at finite temperature is used for  evaluating the dependence of 
  physical quantities such as  the energy density  on the momentum scale  \cite{b14}. 
Furthermore, the holographic renormalization of  gravity 
in asymptotically Lifshitz spacetimes 
naturally reproduces the structure of gravity with anisotropic scaling \cite{c14}. 
The holographic counter-terms induced near anisotropic infinity take the
form of the action for gravity at a Lifshitz point, 
with the appropriate value of the dynamical critical exponent.
The holographic renormalization of Horava-Lifshitz gravity
reproduces the full structure of the $z=2$ anisotropic Weyl anomaly in 
dual field theories in three dimensions \cite{a14}. 
In fact, Lifshitz theories have also become important because of the development of Horava-Lifshitz  gravity
\cite{5}-\cite{9}. 
Horava-Lifshitz  gravity is a renormalizable theory of gravity, in which unitarity is not spoiled. 
Even though gravity is not renormalizable, it can be made renormalizable
by adding higher order curvature terms to it. However, the addition of 
higher order temporal derivatives  spoils the unitarity of the theory. 
A way out of this problem is to 
add higher order spatial derivatives without adding any higher order temporal 
derivatives. 
Even though this  break Lorentz symmetry, the Horava-Lifshitz theory
of gravity reproduces General Relativity in the infrared limit. 

It may be noted that the string theory comes naturally equipped with a minimum measurable length scale, which is the string length scale.
This is because the 
 spacetime  cannot be probed below this scale \cite{string}. 
 Furthermore, the existence of a minimum measurable length scale in a theory produces higher derivative  corrects terms, 
 due to the existence of generalized uncertainty principle \cite{min}. 
 The   the $CFT$ dual to 
 a massive free scalar field theory with such higher derivative corrections has been analysed using the $AdS/CFT$ correspondence \cite{adsf}.  
 So, it will be interesting to analyse the  Lifshitz deformation  of $AdS/CFT$ correspondence, consistent with generalized uncertainty 
 principle.

As we have both the fermionic and bosonic 
actions, it
 will be interesting to analyse supersymmetric theories based on such deformations.
 This can be done by constructing various   supersymmetric theories based on generalized uncertainty principle 
 with Lifshitz scaling. It may be noted that Lifshitz supersymmetric theories 
 have already been constructed \cite{12de}. 
 In fact, 
according to $AdS/CFT$ correspondence  type $IIB$ superstring on $AdS_5\times S^5$ is dual to the maximally
 supersymmetric  $\mathcal{N}=4$ super-Yang-Mills  theory in four dimensions 
\cite{14a}-\cite{17a}, so, this result will can be used to study 
the gravity dual to such a theory.
Furthermore, as  
$
AdS_5\times S^5 \sim SO(2,4)/SO(1,4)\times
SO(6)/SO(5) \subset SU(2,2|4)/SO(1,4)\times SO(5)
$, so, 
the superisometries of this background are generated by the supergroup  $SU(2,2|4)$, which also  
 generates the  superconformal invariance of
$\mathcal{N}=4$ super-Yang-Mills  theory in four dimensions. Here the four dimensional superconformal transformations are  generated by   $SO(2,4)$ 
and the  $R$-symmetry is generated by 
 $SO(6) \sim SU(4)$. Furthermore,  $\mathcal{N} = 4$ super-Yang-Mills theory, with $U(N)$ as the gauge group,
is the low-energy limit for a  stack of multiple coincident D3-branes
on $AdS_5\times S^5$. Here the transverse D3-brane coordinates give rise to six scalar fields 
in the $\mathcal{N} = 4$ super-Yang-Mills theory. Apart from these six bosons, there are also 
sixteen fermions. Thus, a Lifshitz deformation of bulk theory, consistent with generalized uncertainty,  
may produce interesting deformation of the 
$\mathcal{N}=4$ super-Yang-Mills  theory.  
It will be interesting to analyse   such a deformed super-Yang-Mills theories.

\section*{Acknowledgement}
We would like to thank  Ali Nassar for pointing out to us an interesting technique used in conformal field theories, i.e., the
parameters in a conformal field theory
can be promoted to background fields. These background fields can have interesting scaling properties.


\begin{thebibliography}{99}
\bibitem{unz2}D. Amati, M. Ciafaloni and G. Veneziano, Phys. Lett. B 216, 41 (1989)
\bibitem{unzasaqsw}  A. Kempf, G. Mangano, and R. B. Mann, Phys. Rev. D 52, 1108 (1995) 
\bibitem{uncsdcas}L.  N.  Chang, D.  Minic, N. Okamura, and T.  Takeuchi, Phys.Rev. D65,  125027 (2002)
\bibitem{uncscds}L.  N.  Chang, D.  Minic, N. Okamura, and T.  Takeuchi, Phys. Rev. D65, 125028 (2002) 
\bibitem{un2z}S. Benczik, L. N.  Chang, D.  Minic, N.  Okamura, S.  Rayyan, and T.  Takeuchi,  Phys. Rev. D66, 026003 (2002)
\bibitem{unz1}P.  Dzierzak, J.  Jezierski, P.  Malkiewicz, and W. Piechocki,  Acta Phys. Polon. B41, 717 (2010) 
\bibitem{un1}     D. Amati, M. Ciafaloni, and G. Veneziano, Phys. Lett. B  216, 41  (1989) 
\bibitem{un11}    M. Maggiore, Phys. Lett. B 304, 65 (1993)
\bibitem{un12}    M. Maggiore,  Phys. Rev. D   49, 5182  (1994)
\bibitem{un13}    M. Maggiore, Phys. Lett. B   319,  83 (1993)
\bibitem{un14}    L. J. Garay,  Int. J. Mod. Phys. A   10,  145 (1995)
\bibitem{un15}    F. Scardigli, Phys. Lett. B   452, 39  (1999)
\bibitem{un17}    C. Bambi, F. R. Urban,  Class. Quantum Grav.   25,  095006 (2008)
\bibitem{un18}    K. Nozari,  Phys. Lett. B.   629,  41 (2005) 
\bibitem{un19}    K. Nozari, T. Azizi,  Gen. Relativ. Gravit.   38, 735  (2006)
\bibitem{un10}    P. Pedram,  Int. J. Mod. Phys. D   19,  2003 (2010)
\bibitem{un5}     A. Kempf, G. Mangano, and R. B. Mann,  Phys. Rev. D   52,  1108 (1995)
\bibitem{un51}    A. Kempf, J. Phys. A   30, 2093 (1997) 
\bibitem{un52}    F. Brau,  J. Phys. A   32, 7691 (1999) 
\bibitem{un53}    K. Nozari, and B. Fazlpour, Chaos, Solitons and Fractals,   34, 224 (2007) 
\bibitem{un54}    S. Das, and E. C. Vagenas,  Phys. Rev. Lett.   101,  221301 (2008)
\bibitem{unn9}M.  Kober, Phys. Rev. D 82, 085017 (2010)
\bibitem{unskdj}V. Husain, D. Kothawala and S. S. Seahra, Phys. Rev. D 87, 025014 (2013) 
\bibitem{unn}M. Kober, Int. J. Mod. Phys. A 26,   4251 (2011)
\bibitem{1}R. M. Hornreich, M. Luban and S. Shtrikman, , Phys. Rev. Lett. 35, 1678 (1975)
\bibitem{2}G. Grinstein, Phys. Rev. B 23, 4615 (1981)
\bibitem{3}P. M. Chaikin and T. C. Lubensky, Principles of Condensed Matter Physics,
Cambridge University Press, Cambridge, UK (1995)
\bibitem{4}S. Sachdev, Quantum Phase Transitions,  Cambridge University Press, Cambridge, UK (2001)
\bibitem{15}A. Benlagra and M. Vojta, Phys. Rev. B 87, 165143 (2013) 
\bibitem{16}M. Bercx and F. F. Assaad, Phys. Rev. B 86, 075108 (2012) 
\bibitem{a1}P. Gegenwart, Nature Phys. 4, 186 (2008)
\bibitem{a12}P. Coleman, C. Pepin, Q. Si and R. Ramazashvili, J. Phys. Condens. Matt. 13, R723 (2001)
\bibitem{a2}A. Hackl and  M. Vojta, Phys. Rev. Lett. 106, 137002 (2011)
\bibitem{a3}V. B. Svetovoy, Phys. Rev. Lett. 101, 163603 (2008)
\bibitem{a4} S. A. Ellingsen, I. Brevik, J. S. Hoye and K. A. Milton, Phys. Rev. E 78, 021117 (2008)
\bibitem{a5} M. Bordag, B. Geyer, G. L. Klimchitskaya and V. M. Mostepanenko, Phys. Rev. B 74, 205431 (2006)
\bibitem{17}  M.  Alishahiha, M. R.  M.  Mozaffar and A.  Mollabashi,  Phys. Rev. D 86, 026002 (2012) 
\bibitem{18} M.  Liu, J. Lu and J. Lu,   Class. Quant. Grav. 28, 125024 (2011) 
\bibitem{6a}H. Montani and F. A. Schaposnik, Phys. Rev. D 86, 065024 (2012)
\bibitem{7a}L. Caffarelli and L. Silvestre, Comm. Part. Diff. Eqs. 32, 1245 (2007)
\bibitem{8a}R. T. Seeley, Proc. Symp. Pure Math. 10, 288 (1967)
\bibitem{9a}C. Laemmerzahl, J. Math. Phys. 34, 3918 (1993)
\bibitem{10a}J. J. Giambiagi, Nuovo Cim. A 104, 1841 (1991)
\bibitem{12a}C. G. Bollini and J. J. Giambiagi, J. Math. Phys. 34, 610 (1993)
\bibitem{2a}D. Anselmi and M. Halat, Phys. Rev. D 76, 125011 (2007)
\bibitem{3a}D. Anselmi, Eur. Phys. J. C 65, 523 (2010) 
\bibitem{5a}A.  Dhar, G. Mandal and S. R. Wadia, Phys. Rev. D 80, 105018 (2009)
\bibitem{aqwe} Z.  Komargodski, JHEP. 1207, 069 (2012) 
\bibitem{aqwer}Z.  Komargodski, JHEP 1112, 099 (2011) 
\bibitem{5a01} J. Tan, Calc. Var 42, 21 (2011)
 \bibitem{stochastic} G. Parisi and Y. S. Wu, Sci. Sin. 24, 483 (1981)
 \bibitem{csdcds} A. Barchielli and G. Lupieri,  J. Math. Phys. 26,  2222 (1985) 
 \bibitem{dvcs} F. Guerra  and  R. Marra,   Phys. Rev. D29, 1647 (1984) 
 \bibitem{stochastic1}P. H. Damgaard and K. Tsokos, Nucl. Phys. B 235, 75 (1983)
 \bibitem{bh}  P. H. Damgaard and H. Huffel, Phys. Rep. 152, 227 (1987)
 \bibitem{b11h} G. Parisi and N. Sourlas, Nucl. Phys. B 206,  321 (1982) 
 \bibitem{ko} L.  Henriet, Z.  Ristivojevic, P.  P. Orth and  K.  Le Hur,  Phys. Rev. A 90, 023820 (2014)
  \bibitem{11st}U. Lindstrom, M. Rocek, W. Siegel, P. van Nieuwenhuizen, and A. E.
van de Ven, J. Math. Phys. 31, 1761 (1990)
\bibitem{12st}G. Parisi and N. Sourlas, Nucl. Phys. B 206, 321 (1982)
\bibitem{10st}S. Chaturvedi, A. K. Kapoor and V. Srinivasan, Phys. Lett. B 140, 56 (1984)
\bibitem{ferm}H. Montani and F. A. Schaposnik, Phys. Rev. D86, 065024 (2012)
\bibitem{line1} A. F. Ali, S.  Das and E. C. Vagenas,  Phys. Rev. D84, 044013 (2011) 
\bibitem{line2}P. Pedram, K. Nozari and S. H. Taheri,  JHEP 1103, 093 (2011) 
\bibitem{line0}M. Asghari, P. Pedram and K. Nozari, Phys. Lett. B 725, 451  (2013) 
\bibitem{l1}S. Das, E. C. Vagenas and A. F. Ali, Phys. Lett. B 690, 407 (2010) 
\bibitem{l2}W. Chemissany, S. Das, A. F. Ali and  E. C. Vagenas, JCAP. 1112, 017 (2011) 
 \bibitem{13a}J. M. Maldacena, Adv. Theor. Math. Phys. 2, 231 (1998) 
\bibitem{10}S. Kachru, X. Liu and M. Mulligan, Phys. Rev. D78,  106005 (2008)
\bibitem{12}K. Balasubramanian and K. Narayan,  JHEP.  1008, 014 (2010) 
\bibitem{13} R. Gregory, S. L. Parameswaran, G. Tasinato and I. Zavala, JHEP. 1012, 047 (2010) 
\bibitem{14}A. Donos and J. P. Gauntlett,  JHEP. 1012, 002 (2010) 
\bibitem{b14} M. Park and R. B. Mann,  JHEP.  1207, 173 (2012) 
\bibitem{c14}T. Griffin, P. Horava and C. M. M. Thompson, JHEP. 1205, 010 (2012) 
\bibitem{a14}T. Griffin, P. Horava and C. M. M. Thompson, Phys. Rev. Lett. 110,  081602  (2013)
\bibitem{5}P. Horava, Phys. Lett. B 694,172 (2010)
\bibitem{6}P. Horava, Phys. Rev. D 79, 084008 (2009)
\bibitem{7}P. Horava, JHEP. 03, 020 (2009)
\bibitem{8}O. Obregon and J. A. Preciado, Phys. Rev. D 86, 063502 (2012) 
\bibitem{9}A. Sheykhi, Phys. Rev. D 87, 024022 (2013) 
\bibitem{string}D. Amati, M. Ciafaloni and G. Veneziano, Phys. Lett. B 216, 41 (1989)
\bibitem{min}S. Das and E. C. Vagenas, Phys. Rev. Lett. 101, 221301 (2008)
\bibitem{adsf}M. Faizal. A. F. Ali and A. Nassar,  arXiv:1405.4519 (2014)
\bibitem{12de} M. Pospelov and C. Tamarit,   JHEP 01, 048  (2014)
\bibitem{14a}O. Aharony, S. S. Gubser, J. M. Maldacena, H. Ooguri and Y. Oz,  Phys. Rept. 323, 183 (2000)
\bibitem{15a} E. Witten, Adv. Theor. Math. Phys. 2, 253 (1998) 
\bibitem{16a}J. M. Maldacena and C. Nunez, Phys. Rev. Lett. 86, 588 (2001) 
\bibitem{17a}A. Karch and E. Katz, JHEP.  0206, 043 (2002) 
 
\end{thebibliography}
\end{document}